\documentstyle[12pt]{article}

\begin{document}

\begin{titlepage}

\baselineskip=12pt

\begin{flushright}
NEIP-98-004\\
gr-qc/9808046\\
Mod.Phys.Lett.A13:1319-1325,1998
\end{flushright}

\begin{centering}
\vspace{.1in}

{\large {\bf Dimensionful deformations of Poincar\'e 
symmetries}}\\
{\large {\bf for a Quantum Gravity without ideal 
observers}\footnote{This essay received an ``honorable mention'' from 
the Gravity Research Foundation, 1998 --- Ed.}} \\

\vspace{.5in}
{\bf 
Giovanni AMELINO-CAMELIA} \\

\vspace{.2in}

Institut de Physique, Universit\'e de Neuch\^atel,
rue Breguet 1, Neuch\^atel, Switzerland\\
\vspace{.3in}
\vspace{.1in}
{\bf Abstract}\\
\end{centering}

\vspace{.05in}

{\small
Quantum Mechanics is revisited as
the appropriate theoretical framework for
the description of the outcome of experiments that
rely on the use of classical devices.
In particular, it is emphasized that the limitations on the measurability 
of (pairs of conjugate) observables encoded in the
formalism of Quantum Mechanics 
reproduce faithfully the ``classical-device limit'' of the
corresponding limitations
encountered in ({\it real} or {\it gedanken}) experimental setups.
It is then argued that devices cannot behave classically
in Quantum Gravity, and that this 
might raise serious problems for
the search of a class of experiments described by
theories obtained by ``applying Quantum Mechanics to Gravity.''
It is also observed that using heuristic/intuitive arguments 
based on the absence of classical devices 
one is led to consider
some candidate Quantum-Gravity phenomena involving
dimensionful deformations of the Poincar\'e symmetries.}

\end{titlepage}

\newpage

\baselineskip 12pt plus .5pt minus .5pt

\pagenumbering{arabic}
\setcounter{page}{1}
\pagestyle{plain} 

A large portion of the efforts devoted to the
search of a theory encompassing Gravity and Quantum Mechanics
has been focused on the formal development of theories
obtained by ``applying Quantum Mechanics to Gravity.''
The important problem of relating Quantum Gravities obtained in
this way to the ``physical reality'' has been mostly postponed,
but a few studies have gone as far as considering
some related problems such as the ``physical interpretation''
of certain elements emerging in these theories 
({\it e.g.} the physical interpretation of quantum spacetime)
and the identification of a set of observables
for these theories (see {\it e.g.} Ref.~\cite{rovobs}).

The present essay has three objectives.
Firstly, it shall be argued
that analyses of the relation between a 
candidate Quantum Gravity 
and the ``physical reality'' should focus on the search
of a class of experiments that could be described
by such a theory.\footnote{\tenrm
There is a subtle but significant difference
between this position and the one of searching for
the ``physical interpretation'' of the structures
that arise in Quantum Gravity, 
{\it e.g.}, within a given theoretical framework, the search 
of procedures for the measurement of time,
which is natural in the former approach,
might correspond in the latter approach to
the search of a formal structure
to be identified with time.}
In order to clarify which type of relation 
between a theoretical framework and a
counterpart class of experiments 
could be 
desirable,
the familiar case of ordinary
(non-gravitational) Quantum Mechanics and its class of experiments
is here revisited,
emphasizing the role played by ``classical devices.''

The second objective of this essay is to clarify that
devices cannot behave classically
in Quantum Gravity, and that this appears to raise problems
for the search of a class of experiments that could be
described by the above-mentioned theories
obtained by ``applying Quantum Mechanics to Gravity.''

The third and final objective is the one 
of outlining a new theoretical framework for Quantum Gravity
that might be better suited to describe
the outcome of experiments that do not rely 
on ``classical devices.''
While this author is still unable to 
provide a fully developed formalism supporting
this new framework,
certain phenomena that could characterize it
can be discussed on rather general grounds.
In particular, 
they involve dimensionful deformations of Poincar\'e symmetries.

\vskip .19in

\noindent
{\bf Quantum Mechanics and its class of experiments.}
One of the proposals of this essay is that analyses of 
the relation between a candidate Quantum Gravity 
and the ``physical reality'' should focus on the search
of a class of experiments that could be described
by such a theory.
The familiar example of ordinary Quantum Mechanics,
as the appropriate theoretical framework
for the description of the outcome of measurements
performed by ``classical devices'' on a ``quantum system,''
can be used to illustrate
the type of relation that one might expect
between a theoretical framework and a
corresponding class of experiments.
Some intuition can already be gained by just
considering how the limitations on the measurability 
of (pairs of conjugate) observables encoded in the formalism of 
Quantum Mechanics are reflected in the ``classical-device limit'' of the
corresponding limitations encountered in 
the analysis of some ({\it real} or {\it gedanken}) 
experimental setups,\footnote{\tenrm
Although
some of the issues analyzed in this essay
are commonly studied in works on the {\it Foundations of 
Quantum Mechanics}, the objective is very different.
Here one is just trying to build 
up intuition, as a replacement for the intuition
that one ordinarily gets from experimental data (which are
unfortunately not available in the case of Quantum Gravity).
Accordingly the rigorous 
derivations characterizing typical works on the {\it Foundations of Quantum 
Mechanics} can be replaced by more intuitive/heuristic arguments.
In particular, in the following it shall not be necessary to immerge 
too deeply in the labyrinth of measurement theory.
Still, the reader will notice 
that (for simplicity ?) the underlying viewpoint is the one 
of the ``Copenhagen interpretation.''}
and particularly insightful are those studies
which have focused on the fact that, even in Quantum Mechanics,
any given observable can be 
measured with arbitrarily high accuracy (at the cost of loosing all 
information on a conjugate observable).

A heated debate was devoted (several decades ago) to the
search of an experimental setup that would allow to
measure with total accuracy the electromagnetic field.
One such setup was identified in a famous analysis 
by Bohr and Rosenfeld~\cite{rose}.
They showed that under specific conditions
the, say, $x$ component of the (average over a small but finite
world domain of the) electric field  could be measured 
using a continuous charge distribution,
and that the resulting uncertainty $\delta E_x$ turns
out to be proportional to the ratio 
of total electric charge $Q$ 
to inertial mass $M$ of the charge 
distribution: $\delta E_x \propto Q/M$.
The desired
result $\delta E_x = 0$ is then obtained 
by taking the limit $Q/M \rightarrow 0$.

Another very intuitive study
of measurability in Quantum Mechanics
is Wigner's analysis \cite{wign}
of an experimental setup that allows to
measure distances with total accuracy.
In Ref.~\cite{wign}
the distance $L$ between two bodies is measured
by exchanging a light signal between them.
For conceptual simplicity, one can take one of the two bodies 
to be a clock.
The measurement would then be performed 
by {\it attaching}\footnote{\tenrm
Of course,
for consistency with causality,
in such contexts one assumes devices to be ``attached non-rigidly,''
and, in particular, the relative position
and velocity of their centers of mass continue to satisfy the standard 
uncertainty relations of Quantum Mechanics.} 
a light-gun ({\it i.e.} a device 
capable of sending
a light signal when triggered) and a detector
to the clock,
and {\it attaching} a mirror to the other body. 
By measuring the time $T$ needed by the light signal
for a two-way journey between the bodies one 
also
obtains a 
measurement of  $L$.
Within this setup it is easy to realize that 
$\delta L$ can vanish only if 
all devices used in the measurement behave classically.
One can consider for example the
contribution to $\delta L$ coming from 
the uncertainties that affect the motion
of the device composed by the light-gun and the detector.
Denoting with $x^*$ and $v^*$
the position and the velocity of the center of mass
of this device relative to the position of the clock, 
and assuming that the experimentalists prepare this device
in a state with uncertainties $\delta x^*$ and $\delta v^*$,
one easily finds \cite{wign}
\begin{eqnarray}
\delta L \geq 
\delta x^* + T \delta v^* 
\geq 
\delta x^* 
+ { (M_c + M_d) \over 2 M_c \, M_d } { \hbar T \over \delta x^* } 
~,
\label{dawign}
\end{eqnarray}
where $M_c$ is the mass of 
the clock, $M_d$ is the total mass of the device composed of
the light-gun and the detector, 
and the right-hand-side relation
follows from observing that Heisenberg's {\it Uncertainty Principle} 
implies $\delta x^* \delta v^* \ge \hbar (M_c + M_d)/(2 M_c M_d)$.
Clearly, from (\ref{dawign}) it follows that $\delta L = 0$ can
only be achieved in the ``classical-device limit,''
{\it i.e.} the limit of infinitely large $M_c$ and $M_d$.

\vskip .19in

\noindent
{\bf Non-classical devices for Quantum Gravity and the possibility
of deformed Poincar\'e symmetries.} 
The relation
between some aspects of the formalism of Quantum Mechanics and
the nature of the class of experiments it describes
is proposed by this author as a model (of course, just one of the 
possible models one might want to pursue)
for the relation between formalism and class of experiments
in the Quantum-Gravity context.
In particular, it would seem to be desirable
(economical from the conceptual viewpoint)
to find that the limitations on the measurability 
of observables encoded in the Quantum-Gravity formalism
would reproduce faithfully the corresponding limitations
encountered in Quantum-Gravity experimental setups.
The central observation of the present essay is that any theory 
obtained by ``applying Quantum Mechanics to Gravity''
would not have this desirable property.
In fact, Quantum Mechanics
only has this property if the experimental setups involve
devices that ``behave classically,'' and such devices are not 
consistent with the structure of the gravitational 
interactions.

In order to illustrate how the gravitational interactions affect
the behavior of devices it is useful
to reconsider the Bohr-Rosenfeld and the Wigner setups.
A Bohr-Rosenfeld experiment for the gravitational
field \cite{bergstac}
would have to rely on probes with vanishing ratio
of  ``gravitational charge'' 
to inertial mass,
just like probes with vanishing ratio
of  electric charge to inertial mass
are required in order to measure with total 
accuracy the electric field.
However,
one expects that 
even in the short-distance regime the equivalence principle
would fix to 1 
the ratio of gravitational charge 
versus inertial mass.

Concerning the analysis of the Wigner setup in a
gravitational context, a first observation, which
is also believed to hold in various Quantum-Gravity and
String-Theory
approaches~\cite{padma,venezkonish},
is that besides the uncertainties introduced by the devices 
there should also be a measurement-procedure-independent
contribution $L_{QG}$ to the uncertainty in the measurement
of a distance $L$.
Therefore, relation (\ref{dawign}) is replaced by
\begin{eqnarray}
\delta L \geq L_{QG} +
\delta x^* 
+ { (M_c + M_d) \over 2 M \, M_d } { \hbar T \over \delta x^* } 
~.
\label{dawignQG}
\end{eqnarray}
In most Quantum-Gravity scenarios $L_{QG}$ is identified with
the Planck length, whereas in String Theory $L_{QG}$ is the
string length. This ``minimum length'' and the associated minimum 
uncertainty for the measurement of distances provide already 
a very important modification (possibly associated
to non-locality \cite{alu})
of the conceptual framework of Quantum Mechanics;
however, this modification is by now well
accepted and it is not a central element of the analysis reported
in this essay since it does not follow from the nature of devices
in Quantum Gravity. As emphasized in Ref.~\cite{gacmpla},
even more dramatic 
modifications of 
the measurability of distances follow from the fact that
large values of the masses $M_c$ and $M_d$ necessarily lead
to great distorsions of the geometry, and well before 
the $M_c , M_d \! \rightarrow \! \infty$
limit the Wigner measurement procedure can no longer be
completed. 
[For large enough masses we even expect that ``information
walls'' (the ones of black-hole physics) would form between
the various elements of the measurement procedure.]

Having realized that the classical 
limit $M_c , M_d \! \rightarrow \! \infty$ is not 
viable,\footnote{\tenrm
A rigorous definition of a ``classical device'' is 
beyond the scope of this essay. However, it should be emphasized 
that the experimental setups being here considered require
the devices to be accurately positioned during the time
needed for the measurement, and therefore an ideal/classical
device should be infinitely massive so 
that the experimentalists can prepare it in a state 
with $\delta x \, \delta v \sim \hbar/M \sim 0$.
It is the fact that the infinite-mass limit is not accessible
in a gravitational context that forces one to 
consider only ``non-classical devices.'' 
This observation is not inconsistent with
conventional analyses of decoherence for macroscopic systems;
in fact, in appropriate environments, the behavior of a macroscopic 
device will still be ``closer to classical'' than the behavior of
a microscopic device, although the limit in which a device has exactly 
classical behavior is no longer accessible.}
from Eq.(\ref{dawignQG})
one concludes that (as it happens in 
presence of decoherence effects \cite{karo})
uncertainties grow with 
the time $T$ required by the measurement procedure.
In fact, 
from Eq.(\ref{dawignQG}) one arrives \cite{gacmpla}
at a minimum uncertainty
for the measurement of a distance $L$ of the type
\begin{eqnarray}
minimum \left[ \delta L \right] \, \sim \, L_{QG} 
+ \sqrt{{ c T L_{QG}^*}} \,
\sim \, L_{QG} + \sqrt{L \, L_{QG}^*}
~,
\label{gacup}
\end{eqnarray}
where $L_{QG}^*$ is the Quantum-Gravity scale
that takes into account the above-mentioned limitations
due to the absence of classical devices,
and the relation on the right-hand side follows from
the fact that $T$ is naturally proportional to $L$.
Although the length scales $L_{QG}$ and $L_{QG}^*$ arise as
independent entities in the derivation 
of Eq.(\ref{gacup}), it seems plausible \cite{gacmpla} 
that they coincide up to factors of order 1 
({\it e.g.} $L_{QG} \sim L_{QG}^* \sim L_{Planck}$).

While they are not better than heuristic, 
the considerations that lead to 
Eq.(\ref{gacup}) 
are quite plausible and it is probably legitimate to
attempt to use 
Eq.~(\ref{gacup}) 
as a guiding intuition for
the development of a theoretical framework
for Quantum Gravity that would not require classical devices
for its counterpart class of experiments.
As discussed in detail in Ref.~\cite{gacxt},
the fact that according to Eq.~(\ref{gacup})
the minimum uncertainty on $L$ grows with $L$ suggests
that there be violations of the ordinary Poincar\'e symmetries.
Interestingly, Eq.~(\ref{gacup}) can be rederived
(using arguments completely independent from the ones
reviewed above) in the framework of dimensionful ``$\kappa$''
deformations of the Poincar\'e symmetries~\cite{review}.
Both the structure of the $\kappa$-deformed
dispersion relation for massless particles
\begin{eqnarray}
{\bf p}^2 = {\hbar^2 \over L_{QG}^2} \, \left[ 1 -
e^{E L_{QG}/(\hbar c)}\right]^2
\label{dr}
\end{eqnarray}
and the struture of the $\kappa$-deformed
Minkowski space (which in particular assigns $[x_i,t] = x_i L_{QG}/c$) 
have been shown~\cite{gacxt,gaclukinow} to 
be consistent with Eq.~(\ref{gacup}).
In spite of the fact that they do not directly involve
any sort of geometrodynamics,
one is tempted to consider the possibility~\cite{gacxt}
that $\kappa$-deformations 
of Poincar\'e symmetries might provide an effective description 
of certain Quantum-Gravity effects at length scales well above the
Planck length (where one expects to have the onset of ``virulent''
geometrodynamics associated to quantum effects) but well
below the length scales presently accessible experimentally.

In general, it is not surprising that a formalism in which 
the ``classical-device limit'' is not accessible might involve
novel structures at the level of symmetries. In the example of
the Wigner setup one finds that away
from the ``classical-device limit'' the uncertainties
characterizing the dynamics of the devices remain entangled
with the uncertainties on the observable being measured.
Such an entanglement could affect the symmetry
structure of the outcome of a measurement procedure, and in
fact one finds that the uncertainty (\ref{gacup}) is somewhat
affected by the positions of the devices \cite{gacmpla,gacxt}.

The possibility of deformations of Poincar\'e symmetries
here encountered in analyzing the measurability of distances
actually characterizes also other 
intuitive scenarios for Quantum Gravity.
A deformed dispersion relation 
has emerged in studies of the
quantization of point particles in a
discrete (lattice) spacetime (see, {\it e.g.}, Ref.~\cite{thooft}).
Moreover,
deformations of Poincar\'e symmetries
have been discussed for some Quantum-Gravity scenarios
based on Wheeler's ``foamy Quantum-Gravity vacuum,'' since
such a vacuum might provide a preferred frame.
This appears to
be the case 
in the Quantum-Gravity approach
of Ref.~\cite{emn}, where the foamy vacuum is described in 
String-Theory language, and in that context
it was shown \cite{aemn} that the propagation of massless particles
is characterized by a deformed dispersion
relation consistent with (\ref{dr}) and a bound on the measurability
of distances of the type (\ref{gacup}).

\vskip .19in

\noindent
{\bf Closing remarks.}
Some of the points here made rely on heuristic/intuitive arguments,
and accordingly the resulting criticism of the standard approach
to Quantum Gravity is not being proposed by this author
as a definitive reason of skepticism in that approach.
The point of the present essay is rather that the alternative
theoretical framework here advocated is based on a quite plausible
set of intuitions concerning Quantum Gravity and therefore 
it might 
deserve further investigation even in the present atmosphere
of wide-spread (and well-justified)
excitement for the recent progress in the formal development 
of certain theories\footnote{\tenrm
Both in {\it critical string theory}~\cite{gsw} and
in {\it canonical/loop quantum gravity}~\cite{asht}
there have been very significant new developments.
It is perhaps worth emphasizing that these are indeed 
theories obtained by ``applying Quantum Mechanics 
to Gravity,''
even though the starting point for the quantization
is not exactly the formalism developed by Einstein.
In the stringy approach the point particle is replaced
by a string whereas the other approach 
is based on a novel manner of viewing Einstein's equations.}
obtained by ``applying Quantum Mechanics to Gravity.''

The most urgent (and formidable) problem facing the
new Quantum-Gravity approach here advocated
is the one of finding a formalism that would host the network
of structures associated to the novel measurability bound (\ref{gacup}),
a mechanism of decoherence in the foamy Quantum-Gravity vacuum, 
and a dimensionful deformation of the Poincar\'e symmetries
(perhaps, but not necessarily, of type (\ref{dr})).
Among the reasons of interest in this research program one should
also mention
the possibility that, as emphasized
elsewhere \cite{qgess97},
the new Quantum-Gravity approach might
be useful in the understanding of the origin
of the observed matter-antimatter asymmetry.
Moreover, and perhaps more importantly,
some of the above-mentioned phenomena associated to
dimensionful deformations of Poincar\'e symmetries
could soon be tested \cite{grbgac} by exploiting the recent dramatic 
developments in the physics of gamma-ray bursts \cite{grbnews}.
While it is perhaps not surprising that the first Quantum-Gravity
ideas to be tested experimentally
should be very speculative ones, such as those
discussed in this essay, it is nevertheless significant
that the healthy interplay between ``high energy physics''
and astrophysics is finally ready to provide some sort
of experimental input to those searching for Quantum Gravity.

The analysis reported in this essay could also contribute to a 
shift of emphasis for measurability analyses of conventional 
theories obtained by ``applying Quantum Mechanics to Gravity.''
For example, the measurability studies in Ref.~\cite{venezkonish},
which have led to the {\it Enlarged Uncertainty Principle}
of String Theory, have focused on certain formal
elements of the measurement procedure
without providing a complete analysis of an experimental setup.
In particular, all the uncertainties introduced by 
the devices have been neglected, and it might be important to 
understand how these uncertainties
modify the results of Ref.~\cite{venezkonish}.

\bigskip

\baselineskip 12pt plus .5pt minus .5pt

\end{document}